%% file: main.tex
\documentclass[sigconf,screen]{acmart}

\usepackage[utf8]{inputenc} 
\usepackage[T1]{fontenc}
\usepackage{microtype}

\usepackage{amsmath}%
\usepackage{MnSymbol}%
\usepackage{wasysym}%

\usepackage{booktabs} 
\usepackage{multirow}
\usepackage{tabularx}
\usepackage{tabulary}
\usepackage{url}
\usepackage{graphicx}
\usepackage{soul}
\usepackage{numprint}
\npthousandsep{,}
\usepackage{verbatim}
\usepackage[moderate]{savetrees}

\usepackage{array,booktabs} 
\usepackage{subcaption}
\usepackage{hyperref}
\usepackage{enumitem}
\newcommand{\lmttfont}{\fontfamily{lmtt}\selectfont}

\setcopyright{acmcopyright}

\begin{document}

\title{How Bad Can a Bug Get? An Empirical Analysis of Software Failures in the OpenStack Cloud Computing Platform}

\copyrightyear{2019} 
\acmYear{2019} 
\acmConference[ESEC/FSE '19]{Proceedings of the 27th ACM Joint European Software Engineering Conference and Symposium on the Foundations of Software Engineering}{August 26--30, 2019}{Tallinn, Estonia}
\acmBooktitle{Proceedings of the 27th ACM Joint European Software Engineering Conference and Symposium on the Foundations of Software Engineering (ESEC/FSE '19), August 26--30, 2019, Tallinn, Estonia}
\acmPrice{15.00}
\acmDOI{10.1145/3338906.3338916}
\acmISBN{978-1-4503-5572-8/19/08}


\renewcommand{\shorttitle}{Empirical analysis of software failures in the OpenStack ...}


    

\author{Domenico Cotroneo}
\affiliation{\institution{Federico II University of Naples}\country{Italy}}
\email{cotroneo@unina.it}

\author{Luigi De Simone}
\affiliation{\institution{Federico II University of Naples}\country{Italy}}
\email{luigi.desimone@unina.it}

\author{Pietro Liguori}
\affiliation{\institution{Federico II University of Naples}\country{Italy}}
\email{pietro.liguori@unina.it}

\author{Roberto Natella}
\affiliation{\institution{Federico II University of Naples}\country{Italy}}
\email{roberto.natella@unina.it}

\author{Nematollah Bidokhti}
\affiliation{\institution{Futurewei Technologies, Inc.}\country{USA}}
\email{nbidokht@futurewei.com}

\renewcommand{\shortauthors}{D. Cotroneo, L. De Simone, P. Liguori, R. Natella, N. Bidokhti}

\begin{abstract}
Cloud management systems provide abstractions and APIs for programmatically configuring cloud infrastructures. Unfortunately, residual software bugs in these systems can potentially lead to high-severity failures, such as prolonged outages and data losses. In this paper, we investigate the impact of failures in the context widespread OpenStack cloud management system, by performing fault injection and by analyzing the impact of the resulting failures in terms of fail-stop behavior, failure detection through logging, and failure propagation across components. The analysis points out that most of the failures are not timely detected and notified; moreover, many of these failures can silently propagate over time and through components of the cloud management system, which call for more thorough run-time checks and fault containment.
\end{abstract}



\maketitle

\section{Introduction}
\label{sec:introduction}
\input{intro.tex}

\section{Overview on the research problem}
\label{sec:research_problem}
\input{research_problem.tex}

\section{Methodology}
\label{sec:methodology}
\input{methodology.tex}

\section{Experimental results}
\label{sec:experiments}
\input{experiments.tex}

\section{Related work}
\label{sec:related}
\input{related.tex}

\section{Experimental artifacts}
\label{sec:artifacts}
\input{artifacts.tex}

\section{Conclusion}
\label{sec:conclusion}
\input{conclusion.tex}

\section*{Acknowledgments}
This work has been partially supported by the PRIN 2015 project ``GAUSS'' funded by MIUR (Grant n. 2015KWREMX\_002) and by UniNA and Compagnia di San Paolo in the frame of Programme STAR. We are grateful to Alfonso Di Martino for his help in the early stage of this work.

\bibliographystyle{ACM-Reference-Format}
\bibliography{bibliography}

\end{document}

%% file: intro.tex

Cloud management systems, such as \emph{OpenStack} \cite{OpenStack}, are a fundamental element of cloud computing infrastructures. They provide abstractions and APIs for programmatically creating, destroying and snapshotting virtual machine instances; attaching and detaching volumes and IP addresses; configuring security, network, topology, and load balancing settings; and many other services to cloud infrastructure consumers. 
%
%
It is very difficult to avoid software bugs when implementing such a rich set of services: at the time of writing, the OpenStack project codebase consists of more than 9 million lines of code (LoC) \cite{openhub,stackalytics}, which implies thousands of residual software bugs even under the most optimistic assumptions on the bugs-per-LoC density \cite{mcconnell2004code,tanenbaum2006can}. 
As a result of these bugs, many high-severity failures have been occurring in cloud infrastructures of popular providers, causing outages of several hours and the unrecoverable loss of user data \cite{li2018empirical,musavi2016experience,gunawi2014bugs,gunawi2016does}. 

In order to prevent severe failures, software developers invest efforts in mitigating the consequences of residual bugs. Examples are defensive programming practices, such as assertion checking and logging, to timely detect an incorrect state of the system \cite{lyu2007software,florio2008survey} and for providing to system operators useful information for quick troubleshooting \cite{yuan2012improving,yuan2012conservative,farshchi2018metric}. Another important approach to mitigate failures is to implement fault containment strategies. Examples are \textit{i)} interrupting a service as soon as a failure occurs (i.e., a \emph{fail-stop} behavior), by turning high-severity failures, such as data losses, into lower-severity API exceptions that can be gracefully be handled \cite{candea2003crash,swift2006recovering,oppenheimer2003internet}; \textit{ii)} notifying the cloud management system and operators about the failures through error logs, so that they can diagnose issues and undertake recovery actions, such as restoring a previous state checkpoint or backup \cite{weber2012automatic,fu2016process};  \textit{iii)} separating system components across different domains to prevent cascading failures across components \cite{lee1993faults,arlat2002dependability,herder2009fault}.

In this paper, we aim to empirically analyze the impact of high-severity failures in the context of a large-scale, industry-applied case study, to pave the way for failure mitigation strategies in cloud management systems. 
In particular, we analyze the OpenStack project, which is the basis for many commercial cloud management products \cite{OpenStackProducts} and is widespread among public cloud providers and private users \cite{OpenStackUsers}. Moreover, OpenStack is a representative real-world large software system, which includes several sub-systems for managing instances (Nova), volumes (Cinder), virtual networks (Neutron), etc., and orchestrates them to deliver rich cloud computing services. 

We adopt software fault injection to accelerate the occurrence of failures caused by software bugs \cite{chillarege1996:generation-error-set,voas1997:predicting,natella2016assessing}: our approach deliberately injects bugs in one of the system components and analyzes the reaction of the cloud system  in terms of fail-stop behavior, failure reporting through error logs, and failure propagation across components. 
We based fault injection on information on software bugs reported by OpenStack developers and users \cite{Launchpad}, in order to characterize frequent bug patterns occurring in this project. Then, we performed a large fault injection campaign on the three major subsystems of OpenStack (i.e., Nova, Cinder, and Neutron), for a total of 911 experiments. 
The analysis of fault injections pointed out the impact of the injected bugs on the end-users (e.g., service unavailability and resource inconsistencies) and on the ability of the system to recover and to report about the failure (e.g., the contents of log files, and the error notifications raised by the OpenStack service API). 
Results of the experimental campaign revealed the following  findings:

\begin{itemize}[leftmargin=4mm]

\item In the majority of the experiments (55.8\%), OpenStack failures were not mitigated by a fail-stop behavior, leaving resources in an inconsistent state (e.g., instances were not active, volumes were not attached) unbeknownst to the user;
In the 31.3\% of these failures, the problem was never notified to the user through exceptions; the others were only notified after a long delay (longer than 2 minutes on average). 
This behavior threatens data integrity during the period between the occurrence of the failure and its notification (if any) and hinders failure recovery actions.

\item In a small fraction of the experiments (8.5\%), there was no indication of the failure in the logs. These cases represent a high risk for system operators since they lack clues for understanding the failure and restoring the availability of services and resources;

\item In most of the failures (37.5\%), the injected bugs propagated across several OpenStack components. Indeed, 68.3\% of these failures were notified by a  different component from the injected one. 
Moreover, there were relevant cases of failures that caused subtle residual effects on OpenStack (7.5\%): even after removing the injected bug from OpenStack, cleaning-up all virtual resources, and restarting the workload on a set of new resources, the OpenStack services were still experiencing a failure, that could only be recovered by fully restarting the OpenStack platform and restoring its internal database from a backup.
\end{itemize}

These results point out the risk that failures are not timely detected and notified, and that they can silently propagate through the system. Based on this analysis, we identify a set of directions towards more reliable cloud management system. To support future research in this field, we share an artifact for configuring our fault injection environment inside a virtual machine, and our dataset of failures, which includes the injected faults, the workload, the effects of the failures (both the user-side impact and our own in-depth correctness checks), and the error logs produced by OpenStack. 


In the following, Section \ref{sec:research_problem} elaborates on the research problem;  
Section \ref{sec:methodology} describes our  methodology; Section \ref{sec:experiments} presents experimental results; Section \ref{sec:related} discusses related work;
Section \ref{sec:artifacts} includes links to the artifacts to support future research;
Section \ref{sec:conclusion} concludes the paper.

%% file: research_problem.tex


Mitigating the severity of software failures caused by residual bugs is a relevant issue for high-reliability systems \cite{cotroneo2013combining}, yet it still represents an open research challenge. 
Ideally, in the case that a fault occurs, a service should be able to mask the fault or recover from it in a transparent way to the user, such as, by leveraging redundancy. However, this is often not possible in the case of software bugs. 
Since software bugs are \emph{human} mistakes in the source code, the traditional fault-tolerance strategies for hardware and network faults often do not apply. For example, if a service is broken because of a regression bug, then retrying to execute the service API with the same inputs would result again in a failure; a retrial would only succeed in the case that the software bug is triggered by a transient condition, such as a race condition \cite{gray1986computers,grottke2007fighting,carrozza2013analysis}. If recovery is not possible, the failed operation must be necessarily aborted and the user should be notified \cite{netflix2017hystrix,microsoft2017circuit}, so that the failure can be handled at a higher level of the business logic. For example, the end-user can skip the failed operation, or put on hold the workflow until the bug is fixed. 
If the failure does not immediately generate an exception from the OS or from the programming language run-time, the service may continue its faulty execution until it corrupts in subtle ways the results or the state of resources. Such cases need to be mitigated by architecting the software into small, de-coupled components for fault containment, in order to limit the scope of failure (e.g., the \emph{bulkhead} pattern \cite{netflix2017hystrix,microsoft2017bulkhead}); and by applying defensive programming practices to perform redundant checks on the correctness of a service (e.g., pre- and post-conditions to check that a resource has indeed been allocated or updated). In this way, the system can enforce a \emph{fail-stop} behavior of the service (e.g., interrupting an API call that experiences a failure, and generating an exception), so that it can avoid data corruption and limit the outage to a small part of the system (e.g., an individual service call).

In this work, we study the extent of this problem in the context of a cloud management system. Applying software fault tolerance principles in such a large distributed system is difficult since its design and implementation is a trade-off between several objectives, including performance, backward compatibility, programming convenience, etc., which opens to the possibility of failure propagation beyond fault containment limits. 
We investigate this problem from three perspectives, by addressing the following three perspectives.

\vspace{2pt}
\noindent
$\rhd$ \textbf{In the case that service experiences a failure, is it able to exhibit a fail-stop behavior?} 
If a service request could not be completed because of a failure, the service API should return an exception to inform about the issue. Therefore, we experimentally evaluate whether the service indeed halts on failure and whether the failure is explicitly notified to the user. In the worst case, the service API neither halts nor raises an exception, and the state of resources is inconsistent with respect to what the user is expecting (e.g., a VM instance was not actually created, or is indefinitely in the ``\emph{building}'' state).

\vspace{2pt}
\noindent
$\rhd$ \textbf{Are error reporting mechanisms able to point out the occurrence of a failure?} 
Error logs are a valuable source of information for automated recovery mechanisms and system operators to detect failures and restore service availability; and for developers to investigate the root cause of the failure. However, there can be gaps between failures and log messages. We analyze the cases in which the logs do not record any anomalous event related to a failure, since the software may lack checks to detect the anomalous events.

\vspace{2pt}
\noindent
$\rhd$ \textbf{Are failures propagated across the services of the cloud management system?} 
To mitigate the severity of failures, it is desirable that failure is limited to the specific service API that is affected by a software bug. If the failure impacts other services beyond the buggy one (e.g., the incorrect initialization of a VM instance also causes the failure of subsequent operations on the instance), it is more difficult to identify the root cause of the problem and to recover from the failure. Similarly, the failure may cause lasting effects on the cloud infrastructures (e.g., the virtual resources allocated for a failed instance cannot be reclaimed, or interfere with other resource allocations) that are difficult to debug and to recover from. Therefore, we analyze whether failures can spread across different components of the system, and across several service calls.

%% file: methodology.tex

\begin{figure*}[!htb]
  \centering
  \includegraphics[scale=0.27]{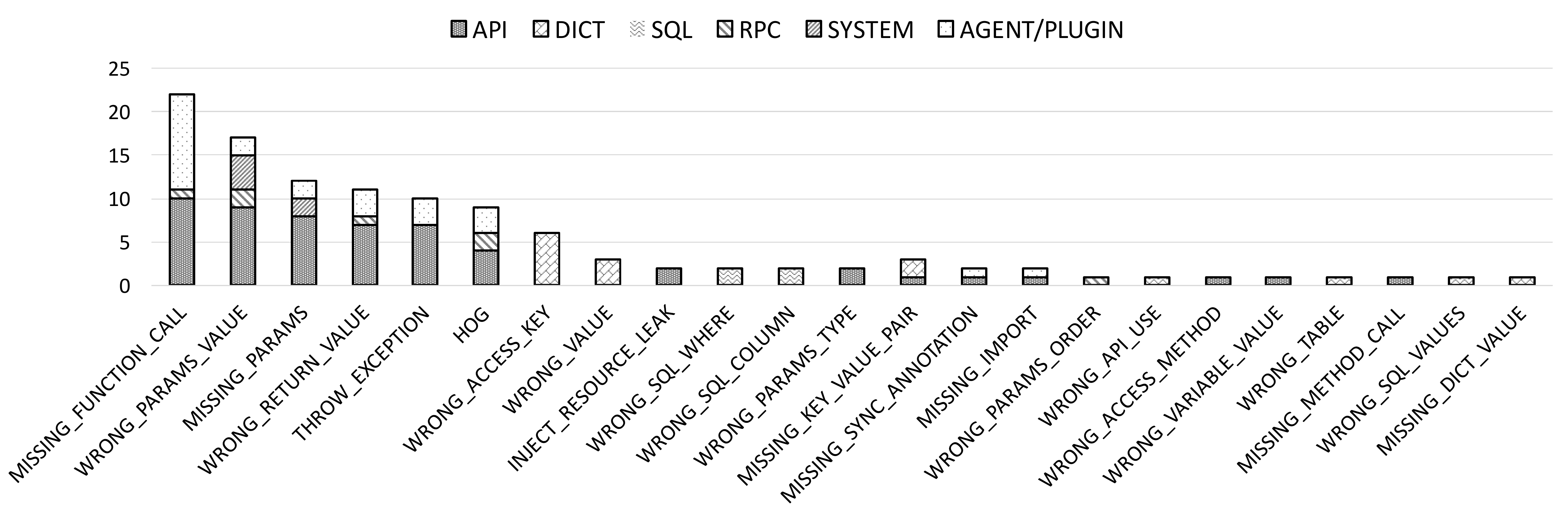}
  \vspace*{-5mm} 
  \caption{Distribution of bug types.}
  \label{fig:faults_nova_neutron}
\vspace{-5.5mm}
\end{figure*}

Our approach is to inject software bugs (\S{}~\ref{subsec:bug_analysis}, \S{}~\ref{subsec:fault_injection}) in order to obtain failure data from OpenStack (\S{}~\ref{subsec:failure_data_collection}). Then, we analyze whether the system could gracefully mitigate the impact of the failures (\S{}~\ref{subsec:failure_analysis}).

\subsection{Bug analysis}
\label{subsec:bug_analysis}







A key aspect to perform software fault injection experiments is to inject representative software bugs \cite{chillarege1996:generation-error-set,duraes2006emulation}. 
Since the body of knowledge on bugs in Python software \cite{rodriguez2018if,orru2015curated}, the programming language of OpenStack, is relatively smaller compared to other languages, we seek for more insights about bugs in the OpenStack project. Therefore, we analyzed the OpenStack issue tracker on the \emph{Launchpad} portal \cite{openstack_launchpad}, by looking for bug-fixes at the source code level, in order to identify \emph{bug patterns} \cite{duraes2006emulation,pan2009toward,martinez2013automatically,zhong2018towards,tufano2018learning} for this project. From these patterns, we defined a set of bug types to be injected.

We went through the problem reports and inspected the related source code. We looked for reports where: (i) the root cause of the problem was a software bug, excluding build, packaging and installation issues; 
(ii) the problem had been marked with the highest severity level (i.e., the problem has a strong impact on  OpenStack services); 
(iii) the problem was fixed, and the bug-fix was linked to the discussion. We manually analyzed a sample of 179 problem reports from the Launchpad, focusing on entries with importance set to ``\emph{Critical}'', and with status set to ``\emph{Fix Committed}'' or ``\emph{Fix Released}'' (such that the problem report also includes a final solution shipped in OpenStack). Of these problem reports, we identified 113 reports that met all of the three criteria. We shared the full set of bug reports (see Section \ref{sec:artifacts}).

The bugs encompass several areas of OpenStack, including: bugs that affected the service APIs exposed to users (e.g., \emph{nova-api}); bugs that affected dictionaries and arrays, such as a wrong key used in {\lmttfont image['imageId']}; bugs that affected SQL queries (e.g., database queries for information about instances in Nova); bugs that affected RPC calls between OpenStack subsystems (e.g., \emph{rpc.cast} was omitted, or had a wrong topic or contents); bugs that affected calls to external system software, such as \emph{iptables} and \emph{dsnmasq}; bugs that affected pluggable modules in OpenStack, such as network protocol plugins and agents in Neutron.
Figure~\ref{fig:faults_nova_neutron} shows statistics about the bug types that we identified from the problem reports and their bug-fixes. The five most frequent bug types include the following ones.


\vspace{2pt}
\noindent
$\blacksquare$ \textbf{Wrong parameters value}: The bug was an incorrect method call inside OpenStack, where a wrong variable was passed to the method call. For example, this was the case of the Nova bug \#1130718 (\url{https://bugs.launchpad.net/nova/+bug/1130718}, which was fixed in \url{https://review.openstack.org/#/c/22431/} by changing the exit codes passed through the parameter {\lmttfont check\_exit\_code}).

\vspace{2pt}
\noindent
$\blacksquare$ \textbf{Missing parameters}: A method call was invoked with omitted parameters (e.g., the method used a default parameter instead of the correct one). For example, this was the case of the Nova bug \#1061166 (\url{https://bugs.launchpad.net/nova/+bug/1061166}, which was fixed in \url{https://review.openstack.org/#/c/14240/} by adding the parameter {\lmttfont read\_deleted='yes'} when calling the SQL Alchemy APIs).

\vspace{2pt}
\noindent
$\blacksquare$ \textbf{Missing function call}: A method call was entirely omitted. For example, this was the case of the Nova bug \#1039400 (\url{https://bugs.launchpad.net/nova/+bug/1039400}, which was fixed in \url{https://review.openstack.org/#/c/12173/} by adding and calling the new method 

\noindent
{\lmttfont trigger\_security\_group\_members\_refresh}).

\vspace{2pt}
\noindent
$\blacksquare$ \textbf{Wrong return value}: A method returned an incorrect value (e.g., {\lmttfont None} instead of a Python object). For example, this was the case of the Nova bug \#855030 (\url{https://bugs.launchpad.net/nova/+bug/855030}, which was fixed in \url{https://review.openstack.org/#/c/1930/} by returning an object allocated through {\lmttfont allocate\_fixed\_ip}).

\vspace{2pt}
\noindent
$\blacksquare$ \textbf{Missing exception handlers}: A method call lacks exception handling. For example, this was the case of the Nova bug \#1096722 ({\url{https://bugs.launchpad.net/nova/+bug/1096722}}, which was fixed in \url{https://review.openstack.org/#/c/19069/} by adding an exception handler for {\lmttfont exception.InstanceNotFound}).


\subsection{Fault injection}
\label{subsec:fault_injection}

\begin{figure*}[!ht]
    \begin{centering}
        \includegraphics[width=0.8\textwidth]{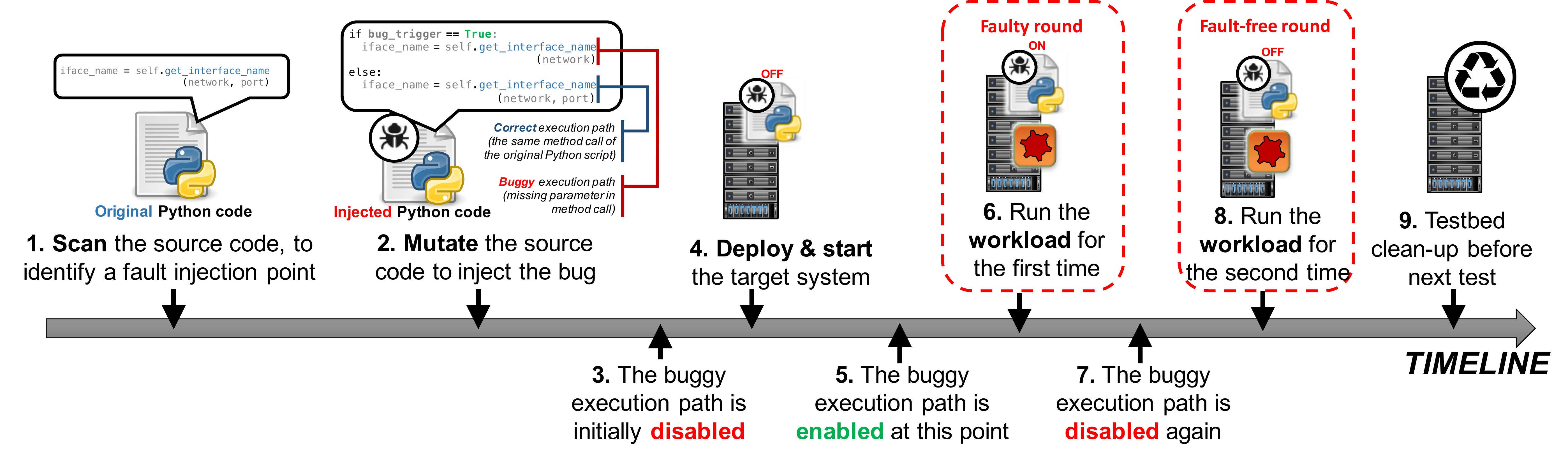}
    \end{centering}
    \vspace*{-5mm} 
    \caption{Overview of a fault injection experiment}
    \label{fig:workflow}
\vspace{-5mm}
\end{figure*}

In this study, we perform \emph{software fault injection} to analyze the impact of software bugs \cite{voas1997:predicting,chillarege1996:generation-error-set,natella2016assessing}. This approach deliberately introduces programming mistakes in the source code, by replacing parts of the original source code with faulty code. 
For example, in \figurename{}~\ref{fig:workflow}, the injected bug emulates a missing optional parameter (a port number) to a function call, which may cause failure under certain conditions (e.g., a VM instance may not be reachable through an intended port).
This approach is based on previous empirical studies, which observed that the injection of code changes can realistically emulate software faults \cite{daran1996software,chillarege1996:generation-error-set,andrews2005mutation}, in the sense that \emph{code changes produce run-time errors that are similar to the ones produced by real software faults}. 
This approach is motivated by the high efforts that would be needed for experimenting with hand-crafted bugs or with real past bugs: in these cases, every bug would require to carefully craft the specific conditions that trigger it (i.e., the topology of the infrastructure, the software configuration, and the hardware devices under which the bug surfaces). 
To achieve a match between injected and real bugs, we focus the injection on the most frequent five types found by the bug analysis. These bug types allow us to cover all of the main areas of OpenStack (API, SQL, etc.), and suffice to generate a large and diverse set of faults over the codebase of OpenStack.

We emulate the bug types by mutating the existing code of OpenStack. 
The \figurename{}~\ref{fig:workflow} shows the steps of a fault injection experiment. 
We developed a tool to automate the bug injection process in Python code. The tool uses the \emph{ast} Python module to generate an \emph{abstract syntax tree} (AST) representation of the source code; then, it scans the AST by looking for relevant elements (function calls, expressions, etc.) where the bug types could be injected; it modifies the AST, by removing or replacing the nodes to introduce the bug; finally, it rewrites the modified AST into Python code, using the \emph{astunparse} Python module. 
To inject the bug types of Section~\ref{subsec:fault_injection}, we modify or remove method calls and their parameters. We targeted method calls related to the bugs that we analyzed, by targeting calls to internal APIs for managing instances, volumes, and networks (e.g., which are denoted by specific keywords, such as \emph{instance} and \emph{nova} for the methods of the Nova subsystem). Wrong input and parameters are injected by wrapping the target expression into a function call, which returns at run-time a corrupted version of the expression based on its data type (e.g., a null reference in place of an object reference, or a negative value in place of an integer). Exceptions are raised on method calls according to a pre-defined list of exception types.

The tool inserts fault-injected statements into an \emph{if} block, together with the original version of the same statements but in a different branch (as in step 2 in \figurename{}~\ref{fig:workflow}). 
The execution of the fault-injected code is controlled by a \emph{trigger} variable, which is stored in a shared memory area that is writable from an external program. This approach has been adopted for controlling the occurrence of failures during the tests. In the first phase (\textbf{round 1}), we enable the fault-injected code, and we run a workload that exercises the service APIs of the cloud management system. During this phase, the fault-injected code could generate run-time errors inside the system, which will potentially lead to user-perceived failures. Afterward, in a second phase (\textbf{round 2}), we disable the injected bug, and we execute the workload for a second time. This fault-free execution points out whether the scope of run-time errors (generated by the first phase) is limited to the service API invocations that triggered the buggy code (e.g., the bug only impacts on local session data). If failures still occur during the second phase, then the system has not able to handle the run-time errors of the first phase. Such failures point out the propagation of effects across the cloud management system (see \S~\ref{sec:research_problem}).


We implemented a workload generator to automatically exercise the service APIs of the main OpenStack sub-systems. The workload has been designed to cover several sub-systems of OpenStack and several types of virtual resources, in a similar way to integration test cases from the OpenStack project \cite{openstack_tempest}. The workload creates VM instances, along with key pairs and a security group; attaches the instances to volumes; creates a virtual network, with virtual routers; and assigns floating IPs to connect the instances to the virtual network. Having a comprehensive workload allows us to point out propagation effects across sub-systems caused by bugs.

The experimental workflow is repeated several times. Every experiment injects a different fault, and only one fault is injected per experiment. Before a new experiment, we clean-up any potential residual effect from the previous experiment, in order to be able to relate failure to the specific bug that caused it. The clean-up re-deploys OpenStack removes all temporary files and processes and restores the database to its initial state. However, we do not perform these clean-up operations between the two workload rounds (i.e., no clean-up between the steps 6 and 8 of \figurename{}~\ref{fig:workflow}), since we want to assess the impact of residual side effects caused by the bug.

\begin{table}[t]
    \begin{center}
        \caption{Assertion check failures.}
        \label{tab:assertion_and_description}
        \vspace{-4mm}
        
        {
        \scriptsize

        \begin{tabulary}{\columnwidth}{|L|L|}
        
            \hline
            \textbf{Name} & \textbf{Description}\\
            \hline
            FAILURE IMAGE ACTIVE & The created \textit{image} does not transit into the \textit{ACTIVE} state\\ \hline
            FAILURE INSTANCE ACTIVE & The created \textit{instance} does not transit into the \textit{ACTIVE} state \\ \hline
            FAILURE SSH & It is impossible to establish a \textit{ssh} session to the created instance\\ \hline
            FAILURE KEYPAIR &  The creation of a \textit{keypair} fails\\ \hline
            FAILURE SECURITY GROUP &  The creation of a \textit{security group} and  \textit{rules} fails\\ \hline
            FAILURE VOLUME CREATED &  The creation of a \textit{volume} fails\\ \hline
            FAILURE VOLUME ATTACHED & Attaching a \textit{volume} to an instance fails\\ \hline
            FAILURE FLOATING IP CREATED & The creation of a \textit{floating IP} fails\\ \hline
            FAILURE FLOATING IP ADDED & Adding a \textit{floating IP} to an instance fails\\ \hline
            FAILURE PRIVATE NETWORK ACTIVE & The created \textit{network} resource does not transit into the \textit{ACTIVE} state\\ \hline
            FAILURE PRIVATE SUBNET CREATED & The creation of a \textit{subnet} fails\\ \hline
            FAILURE ROUTER ACTIVE & The created \textit{router} resource does not transit into the \textit{ACTIVE} state\\ \hline
            FAILURE ROUTER INTERFACE CREATED & The creation of a router interface fails\\ \hline

        \end{tabulary}
        }
        
    \end{center}
    
    \vspace{-0.65cm}
    
\end{table}

\subsection{Failure data collection}
\label{subsec:failure_data_collection}

During the execution of the workload, we record inputs and outputs of service API calls of OpenStack. Any exception generated from the call (\emph{API Errors}) is also recorded. In-between calls to service APIs, the workload also performs \emph{assertion checks} on the status of the virtual resources, in order to point out failures of the cloud management system. 
In the context of our methodology, assertion checks serve as \emph{ground truth} about the occurrence of failures during the experiments. These checks are valuable to point out the cases in which a fault causes an error, but the system does not generate an API error (i.e., the system is unaware of the failure state). 
Our assertion checks are similar to the ones performed by the integration tests as test oracles  \cite{ju2013fault,openstack_instances_states}: they assess the connectivity of the instances through SSH and query the OpenStack API to check that the status of the instances, volumes and network is consistent with the expectation of the test cases. 
The assertion checks are performed by our workload generator. For example, after invoking the API for creating a volume, the workload queries the volume status to check if it is available (\emph{VOLUME CREATED assertion}). These checks are useful to find failures not notified through the API errors.
\tablename{}~\ref{tab:assertion_and_description} describes the assertion checks.

If an API call generates an error, the workload is aborted, as no further operation is possible on the resources affected by the failure (e.g., no volume could be attached if the instance could not be created). In the case that the system fails without raising an exception (i.e., an assertion check highlights a failure, but the system does not generate an API error), the workload continues the execution (as a hypothetical end-user, being unaware of the failure, would do), regardless of failed assertion check(s). The workload generator records the outcomes of both the API calls and of the assertion checks. Moreover, we collect all the log files generated by the cloud management system. This data is later analyzed for understanding the behavior of the system under failure.

\subsection{Failure analysis}
\label{subsec:failure_analysis}

We analyze fault injection experiments according to three perspectives discussed in Section~\ref{sec:research_problem}. 
The first perspective classifies the experiments \emph{with respect to the type of failure that the system experiences}. The possible cases are the following ones.


\vspace{2pt}
\noindent
$\blacksquare$ \textbf{API Error}: In these cases, the workload was not able to correctly execute, due to an exception raised by a service API call. In these cases, the cloud management system has been able to handle the failure in a fail-stop way, since the user is informed by the exception that the virtual resources could not be used, and it can perform recovery actions to address the failure. In our experiments, the workload stops on the occurrence of an exception, as discussed before.

\vspace{2pt}
\noindent
$\blacksquare$ \textbf{Assertion failure}: In these cases, the failure was not pointed out by an exception raised by a service API. The failure was detected by the assertion checks made by the workload in-between API calls, which found an incorrect state of virtual resources. In these cases, the execution of the workload was not interrupted, as no exception was raised by the service APIs during the whole experiment, and the service API did (apparently) work from the perspective of the user. These cases point out non-fail-stop behavior.

\vspace{2pt}
\noindent
$\blacksquare$ \textbf{Assertion failure(s), followed by an API Error}: In these cases, the failure was initially detected by assertion checks, which found an incorrect state of virtual resources in-between API calls. After the assertion check detected the failure, the workload continued the execution, by performing further service API calls, until an API error occurred in a later API call. These cases also point out issues at handling the failure, since the user is unaware of the failure state and cannot perform recovery actions.

\vspace{2pt}
\noindent
$\blacksquare$ \textbf{No failure}: The injected bug did not cause a failure that could be perceived by the user (neither by API exceptions nor by assertion checks). It is possible that the effects of the bug were tolerated by the system (e.g., the system switched to an alternative execution path to provide the service); or, the injected source code was harmless (e.g., an uninitialized variable is later assigned before use). Since no failure occurred, these experiments are not further analyzed, as they do not allow to draw conclusions on the failure behavior of the system.

\vspace{2pt}

Failed executions are further classified according to a second perspective, \emph{with respect to the execution round in which the system experienced a failure}. The possible cases are the following ones.


\vspace{1pt}
\noindent
$\rhd$ \textbf{Failure in the faulty round only}: In these cases, a failure occurred in the first (faulty) execution round (\figurename{}~\ref{fig:workflow}), in which a bug has been injected; and no failure is observed during the second (fault-free) execution round, in which the injected bug is disabled, and in which the workload operates on a new set of resources. This behavior is the likely outcome of an experiment since we are deliberately forcing a service failure only in the first round through the injected bug.

\vspace{1pt}
\noindent
$\rhd$ \textbf{Failure in the fault-free round (despite the faulty round)}: 
These cases are concerns for fault containment since the system is still experiencing failures despite the bug is disabled after the first round and the workload operates on a new set of resources. This behavior is due to residual effects of the bug that propagated through session state, persistent data, or other shared resources.


\vspace{1pt}

Finally, the experiments with failures are classified from the perspective of \emph{whether they generated logs able to indicate the failure}. In order to make more resilient a system, we are interested in whether it produces information for detecting failures and for triggering recovery actions. 
In practice, developers are conservative at logging information for post-mortem analysis, by recording high volumes of low-quality log messages that bury the truly important information among many trivial logs of similar severity and contents, making it difficult to locate issues \cite{zhu2015learning,li2017log,yuan2012improving}. Therefore, we cannot simply rely on the presence of logs to conclude that a failure was detected.

\begin{figure}[t]
    \begin{centering}
        \includegraphics[width=0.95\columnwidth]{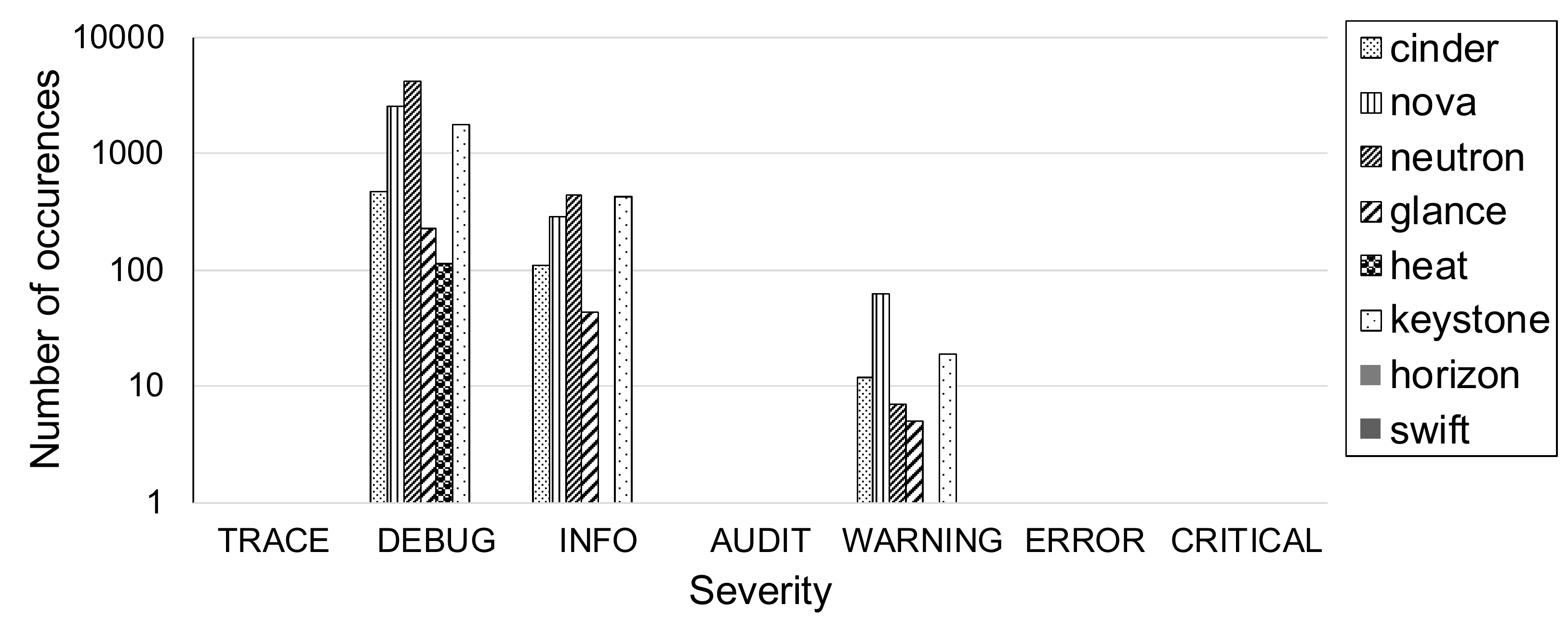}
    \end{centering}
    \vspace*{-5mm} 
    \caption{Distribution of log messages severity during a fault-free execution of the workload.}
    \label{fig:logging_stats}
    \vspace{-6mm}
\end{figure}

To clarify the issue, \figurename{}~\ref{fig:logging_stats} shows the distribution of the number of log messages in OpenStack across severity levels, \textit{TRACE} to \textit{CRITICAL}, during the execution of our workload generator, and \emph{without} any failure. We can notice that all OpenStack components generate a large number of messages with severity \textit{WARNING}, \textit{INFO}, and \textit{DEBUG} even when there is no failure. Instead, there are no messages of severity \textit{ERROR} or \textit{CRITICAL}. Therefore, even if a failure is logged with severity \textit{WARNING} or lower, such log messages cannot be adopted for automated detection and recovery of the failure, as it is difficult to distinguish between ``informative'' messages and actual issues. Therefore, to evaluate the ability of the system to support recovery and troubleshooting through logs, we classify failures according to the presence of one or more \emph{high-severity message} (i.e., \emph{CRITICAL} or \emph{ERROR}) recorded in the log files (\textbf{logged failures}), or no such message (\textbf{non-logged failures}).

%% file: experiments.tex







In this work, we present the analysis of OpenStack version 3.12.1 (release \textit{Pike}), which was the latest version of OpenStack when we started this work. We injected bugs into the most fundamental services of OpenStack \cite{denton2015learning,solberg2017openstack}: (i) the \textbf{Nova} subsystem, which provides services for provisioning instances (VMs) and handling their life cycle; (ii) the \textbf{Cinder} subsystem, which provides services for managing block storage for instances; and (iii) the \textbf{Neutron} subsystem, which provides services for provisioning virtual networks for instances, including resources such as \emph{floating IPs}, \emph{ports} and \emph{subnets}. Each subsystem includes several components (e.g., the Nova sub-system includes \emph{nova-api}, \emph{nova-compute}, etc.), which interact through message queues internally to OpenStack. The Nova, Cinder, and Neutron sub-systems provide external REST API interfaces to cloud users.


%
%
%



\begin{figure}[t]
    \begin{centering}
        \includegraphics[width=0.75\columnwidth]{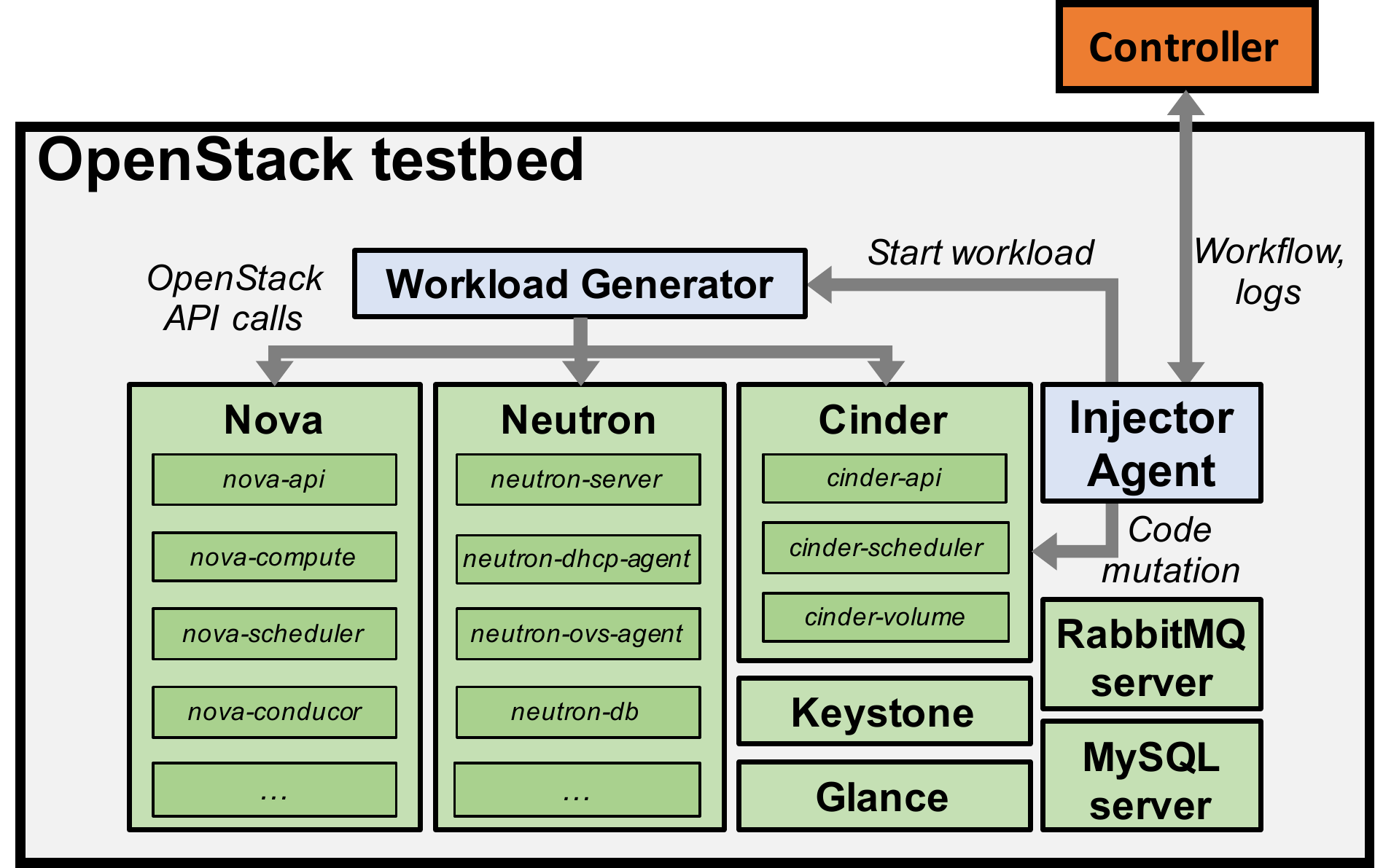}
    \end{centering}
    \vspace*{-4.5mm}
    \caption{OpenStack testbed architecture.}
    \label{fig:testbed_architecture}
\vspace{-0.8cm}
\end{figure}

\figurename{}~\ref{fig:testbed_architecture} shows the testbed used for the experimental analysis of OpenStack. We adopted an all-in-one virtualized deployment of OpenStack, in which the OpenStack services run on the same VM, for the following reasons: (1) to prevent interferences on the tests from transient issues in the physical network (e.g., sporadic network faults, network delays caused by other user traffic in our local data center, etc.); (2) to parallelize a high number of tests on several physical machines, by using the \emph{Packstack} installation utility \cite{packstack} to have a reproducible installation of OpenStack across the VMs; (3) to efficiently revert any persistent effect of a fault injection test on the OpenStack deployment (e.g., file system issues), in order to assure independence among the tests. Moreover, the all-in-one virtualized deployment is a common solution for performing tests on OpenStack \cite{evaluating_openstack,Markelov2016}. The hardware and VM configuration for the testbed includes: 8 virtual Intel Xeon CPUs (E5-2630L v3 @ 1.80GHz); 16GB RAM; 150 GB storage; Linux CentOS v7.0.

In addition to the core services of OpenStack (e.g., Nova, Neutron, Cinder, etc.), the testbed also includes our own components to automate fault injection tests. 
The \emph{Injector Agent} is the component that analyzes and instruments the source code of OpenStack. The \emph{Injector Agent} can: (i) scan the source code to identify injectable locations (i.e., source-code statements where the bug types discussed in \S{}~\ref{subsec:fault_injection} can be applied); (ii) instrument the source code by introducing logging statements in every injectable location, in order to get a profile of which locations are covered during the execution of the workload (\textbf{coverage analysis}); (iii) instrument the source code to introduce a bug into an individual injectable location.

The \emph{Controller} orchestrates the experimental workflow. 
It first commands the \emph{Injector Agent} to perform a preliminary coverage analysis, by instrumenting the source code with logging statements, restarting the OpenStack services, and launching the \emph{Workload Generator}, but without injecting any fault. 
The \emph{Workload Generator} issues a sequence of API calls in order to stimulate OpenStack services.
The \emph{Controller} retrieves the list of injectable locations and their coverage from the \emph{Injector Agent}. Then, it iterates over the list of injectable locations that are covered, and issues commands for the \emph{Injector Agent} to perform fault injection tests. For each test, the \emph{Injector Agent} introduces an individual bug by mutating the source code, restarts the OpenStack services, starts the workload, and triggers the injected bug as discussed in \S{}~\ref{subsec:fault_injection}. The \emph{Injector Agent} collects the logs files from all OpenStack subsystems and from the \emph{Workload Generator}, which are sent to the \emph{Controller} for later analysis (\S{}~\ref{subsec:failure_analysis}).

\begin{figure}[t]
    \begin{centering}
        \includegraphics[width=0.72\columnwidth]{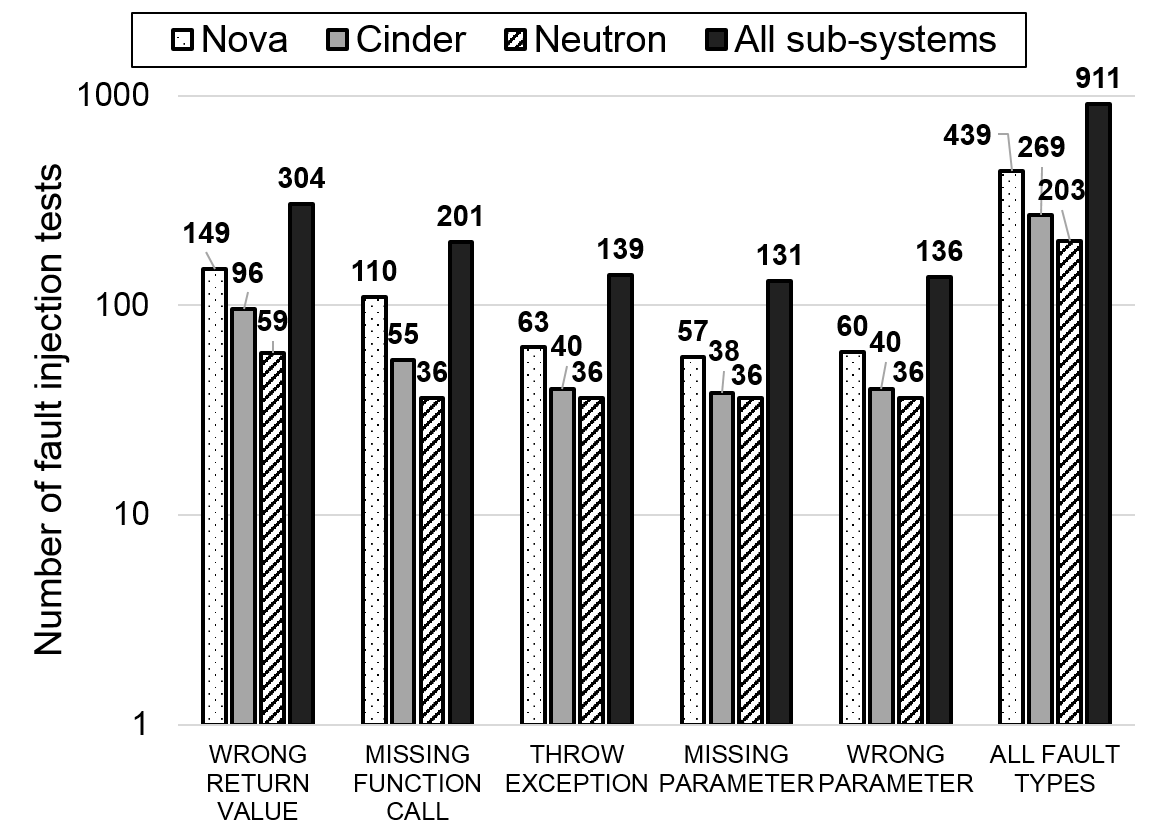}
    \end{centering}
    \vspace*{-5mm} 
    \caption{Number of fault injection tests.}
    \label{fig:fault_injection_tests}
 \vspace{-5.3mm}
\end{figure}

We performed a full scan of injectable locations in the source code of Nova, Cinder, and Neutron, for a total of \numprint{2016} analyzed source code files. We identified \numprint{911} injectable faults that were covered by the workload. \figurename{}~\ref{fig:fault_injection_tests} shows the number of faults per sub-system and per type of fault. 
The number of faults for each type and sub-system depends on the number of calls to the target functions, and on their input and output parameters, as discussed in \S{}~\ref{subsec:fault_injection}.
We executed one the test per injectable location, by injecting one fault at a time. 

\begin{figure}[t]
    \begin{centering}
        \includegraphics[width=0.72\columnwidth]{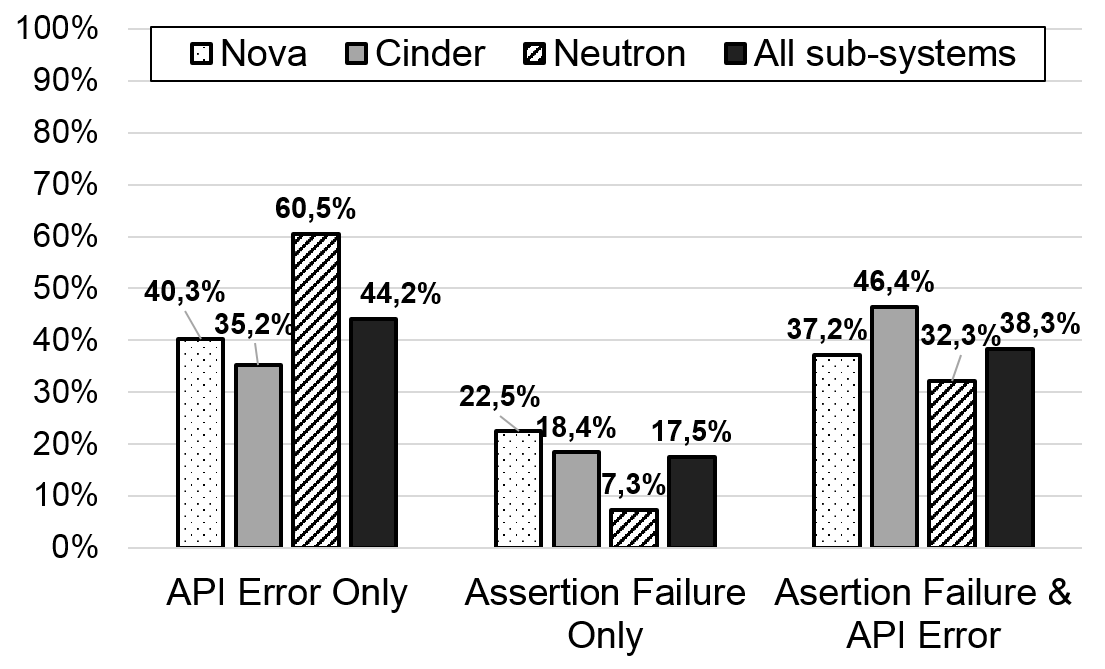}
        \vspace{-0.5cm}
    \end{centering}
    \caption{Distribution of OpenStack failures.}
    \label{fig:all_component_failure_types}
    \vspace{-0.75cm}
\end{figure}

After executing the tests, we found failures respectively in 52.6\% (231 out of 439 tests), 46.4\% (125 out of 269 tests), and 61\% (124 out of 203 tests) of tests in Nova, Cinder, and Neutron, for a total of \numprint{480}.
In the remaining 47.3\% of the tests (431 out of 911 tests), instead, there were neither an API error nor assertion failures: in these cases, the fault was not activated (even if the faulty code was covered by the workload), or there was no error propagation to the component interface. The occurrence of tests not causing failures is a typical phenomenon that occurs with code mutations, which may not infect the state even when the faulty code is executed \cite{christmansson1996generation,lanzaro2014empirical}. Yet, the injections provided us a large and diverse set of failures for our analysis.




\subsection{Does OpenStack show a fail-stop behavior?}
\label{subsec:rq1}

We first analyze the impact of failures on the service interface APIs provided by OpenStack. 
The \emph{Workload Generator} (which impersonates a user of the cloud management system) invokes these APIs, looks for errors returned by the APIs and performs assertion checks between API calls. A fail-stop behavior occurs when an API returns an error before any failed assertion check. In such cases, the \emph{Workload Generator} stops on the occurrence of the API error. Instead, it is possible that an API invocation terminates without returning any error, but leaving the internal resources of the infrastructure (instances, volumes, etc.) in a failed state, which is reported by assertion checks. These cases represent violations of the fail-stop hypothesis, and represent a risk for the users as they are unaware of the failure.
To investigate this aspect, we initially focus on the faulty round of each test, in which fault injection is enabled (\figurename{}~\ref{fig:workflow}). 

\figurename{}~\ref{fig:all_component_failure_types} shows the number of tests that experienced failures, divided into \emph{API Error only}, \emph{Assertion Failure only}, and \emph{Assertion Failure(s), followed by an API Error}. The figure shows the data divided with respect to the subsystem where the bug was injected (respectively in Nova, Cinder, and Neutron); moreover, \figurename{}~\ref{fig:all_component_failure_types} shows the distribution across all fault injection tests. 
We can see that the cases in which the system does not exhibit a fail-stop behavior (i.e., the categories \emph{Assertion Failure only} and \emph{Assertion Failure followed by an API Error})  represent the majority of the failures.

\begin{figure}[t]
    \begin{centering}
        \includegraphics[width=.75\columnwidth]{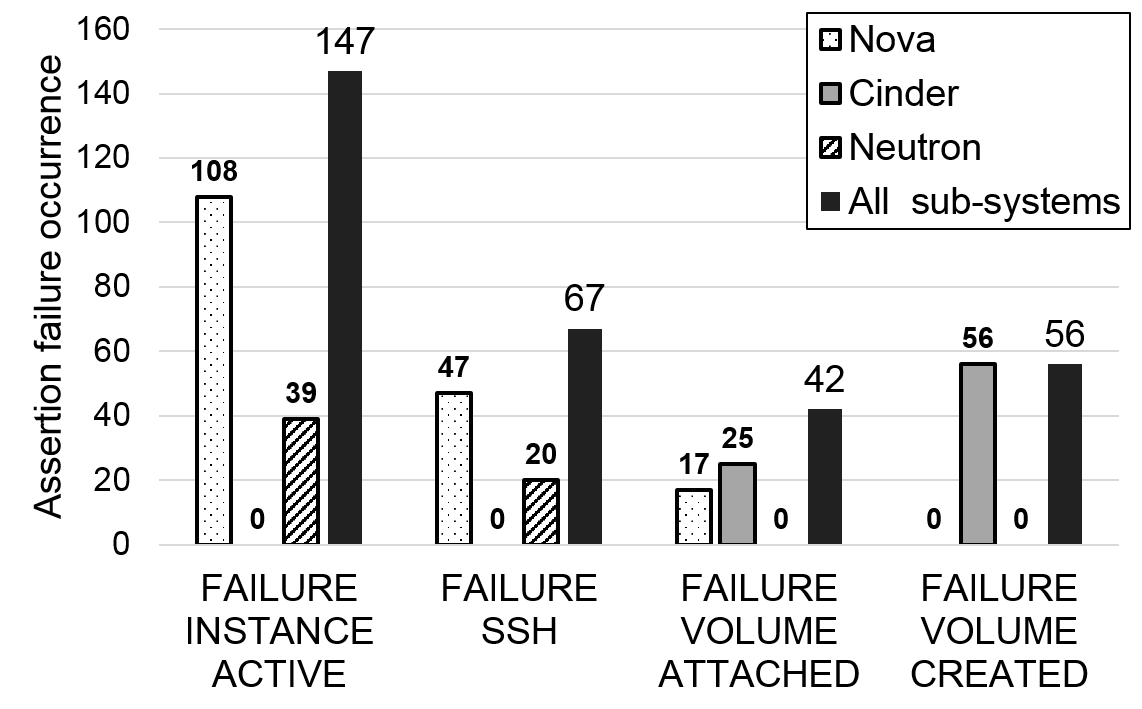}
    \end{centering}
    \vspace*{-5mm} 
    \caption{Distribution of assertion check failures.}
    \label{fig:all_component_failure_assertion_types}
    \vspace{-0.1cm}
\end{figure}

\figurename{}~\ref{fig:all_component_failure_assertion_types} shows a detailed perspective on the failures of assertion checks. Notice that the number of assertion is greater than the number of tests classified in the Assertion failure category (i.e., \emph{Assertion Failure only} and \emph{Assertion Failure followed by an API Error}) since a test can generate multiple assertion failures.
The most common case has been one of the instances not active because the instance creation failed (i.e., it did not move into the \textit{ACTIVE} state \cite{openstack_instances_states}). 
In other cases, the instance could not be reached through the network or could not be attached to a volume, even if in the \textit{ACTIVE} state. A further common case is the failure of the volume creation, but only the faults injected in the Cinder sub-system caused this assertion failure.


\begin{figure}[t]
    \vspace{-0.4cm}
    \begin{centering}
        \includegraphics[width=0.6\columnwidth]{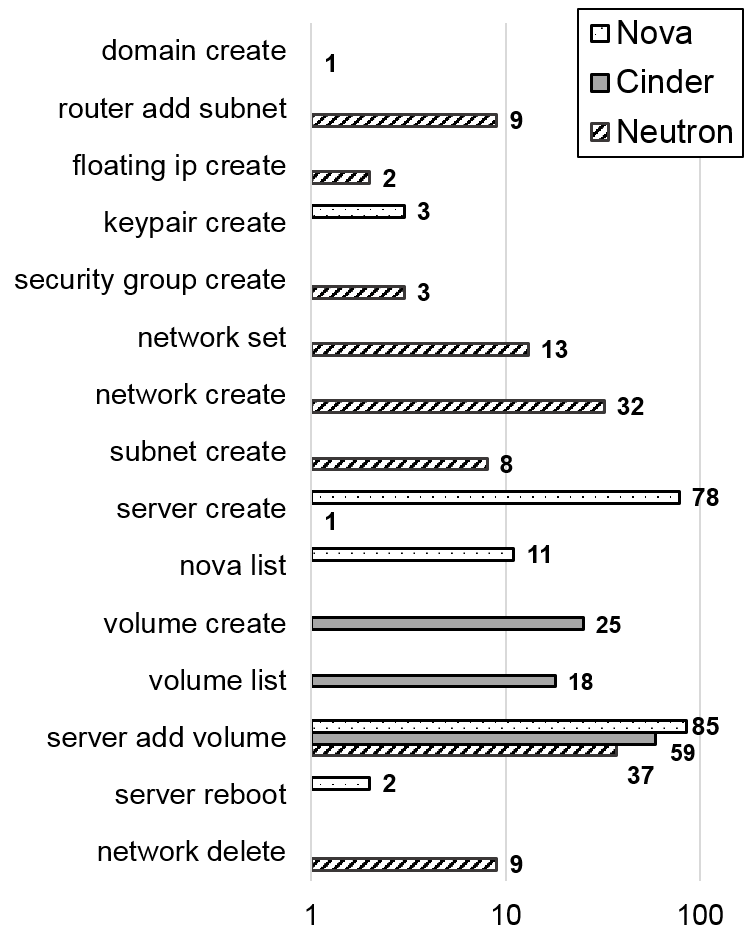}
    \end{centering}
    \vspace*{-5mm} 
    \caption{Distribution of API Errors.}
    \label{fig:all_component_api_errros_after_assertion}
    \vspace{-7mm}
\end{figure}

These cases point out that OpenStack lacks redundant checks to assure that the state of the virtual resources after a service call is in the expected state (e.g., newly-created instances are active). Such redundant checks would assess the state of the virtual resources before and after a service invocation and would raise an error if the state does not comply with the expectation (such as a new instance could not be activated). However, these redundant checks are seldom adopted, most likely due to the performance penalty they would incur, and because of the additional engineering efforts to design and implement them. Nevertheless, the cloud management system is exposed to the risk that residual bugs can lead to non-fail-stop behaviors, where failures are notified with a delay or not notified at all. This makes not trivial to prevent data losses and to automate recovery actions.

\figurename{}~\ref{fig:all_component_api_errros_after_assertion} provides another perspective on API errors. It shows the number of tests in which each API returned an error, focusing on 15 out of 40 APIs that failed at least one time. The API with the highest number of API errors is the one for adding a volume to an instance (\textit{openstack server add volume}), provided by the Cinder sub-system. 
This API generated errors even when faults were injected in Nova (instance management) and Neutron (virtual networking). This behavior means that the effects of fault injection propagated from other sub-systems to Cinder (e.g., if an instance is in an incorrect state, other APIs on that resource are also exposed to failures). On the one hand, this behavior is an opportunity for detecting failures, even if in a later stage. On the other hand, it also represents the possibility of a failure to spread across sub-systems, thus defeating fault containment and exacerbating the severity of the failure. We will analyze fault propagation in more detail in Section~\ref{subsec:rq3}.



\begin{figure}[!t]
    \vspace{-3mm}
    \begin{centering}
        \includegraphics[width=0.75\columnwidth]{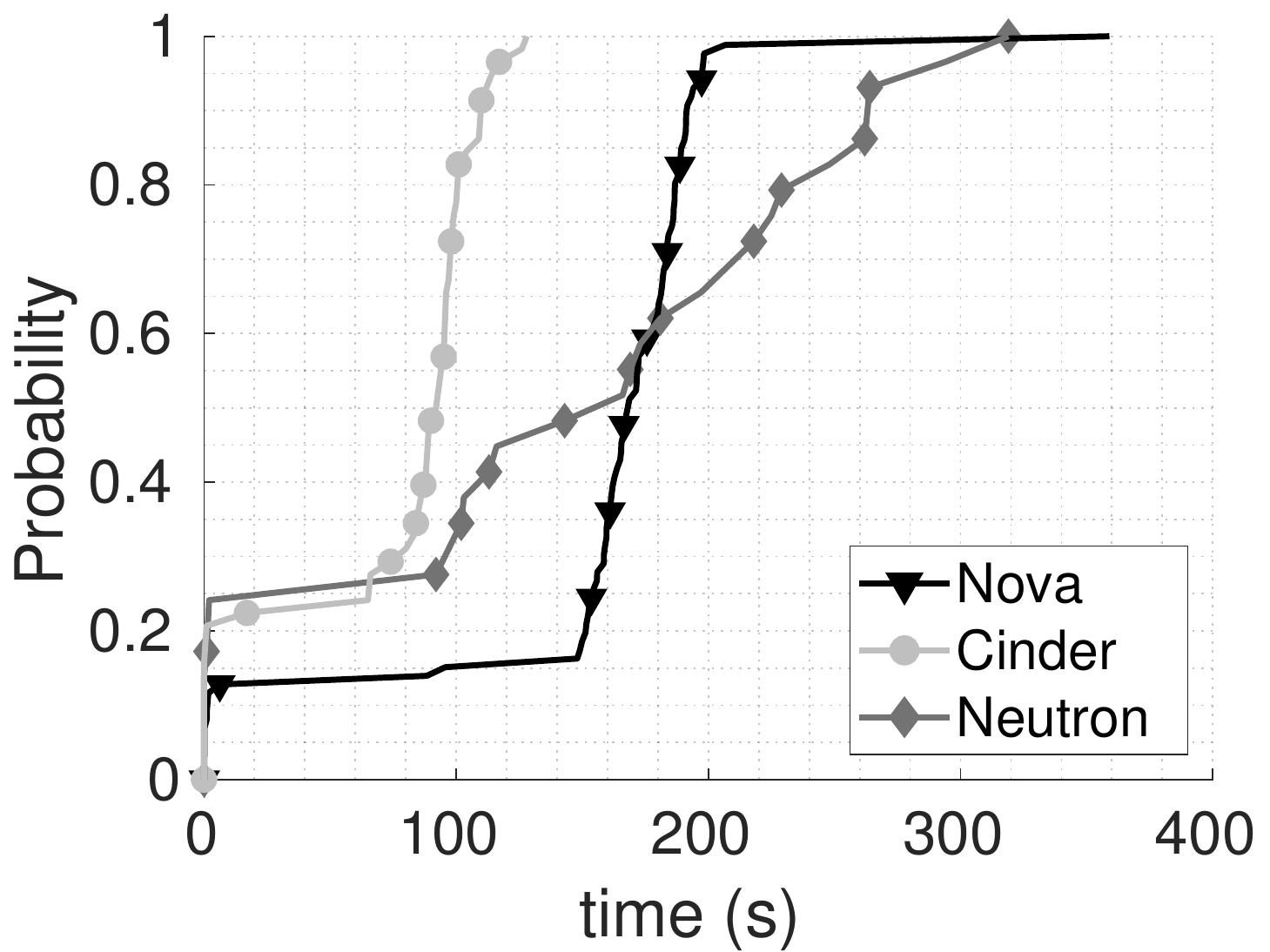}
    \end{centering}
    \vspace*{-5mm} 
    \caption{Cumulative distribution of API Error latency.}
    \label{fig:all_components_assertion_and_api_errors_latency}
    \vspace{-0.92cm}
\end{figure}

To understand the extent of non-fail-stop behaviors, we also analyze the period of time (\textbf{latency}) between the execution of the injected bug and the resulting API error. 
It is desirable that this latency is as low as possible. Otherwise, the longer the latency, the more difficult is to relate an API error with its root cause (i.e., an API call invoked much earlier, on a different sub-system or virtual resource); and the more difficult to perform troubleshooting and recovery actions. 
To track the execution of the injected bug, we instrumented the injected code with logging statements to record the timestamp of its execution. 
If the injected code is executed several times before a failure (e.g., in the body of a loop), we conservatively consider the last timestamp. 
We consider separately the cases where the API error is preceded by assertion check failures (i.e., the injected bug was triggered by an API different from the one affected by the bug) from the cases without any assertion check failure (e.g., the API error arises from the same API affected by the injected bug).

\figurename{}~\ref{fig:all_components_assertion_and_api_errors_latency} shows the distributions of latency for API errors that occurred after assertion check failures, respectively for the injections in Nova, Cinder, and Neutron. Table~\ref{tab:all_components_all_latencies_table_last_activation} summarizes the average, the 50$^{th}$, and the 90$^{th}$ percentiles of the latency distributions. We note that in the first category (API errors after assertion checks), all sub-systems exhibit a median API error latency longer than 100 seconds, with cases longer than several minutes. This latency should be considered too long for cloud services with high-availability SLAs (e.g., four \emph{nines} or more \cite{Bauer:2012:RAC:2339445}), which can only afford few minutes of monthly outage. 
This behavior points out that the API errors are due to a ``reactive'' behavior of OpenStack, which does not actively perform any redundant check on the integrity of virtual resources, but only reacts to the inconsistent state of the resources once they are requested in a later service invocation. Thus, OpenStack experiences a long API error latency when a bug leaves a virtual resource in an inconsistent state. 
This result suggests the need for improved error checking mechanisms inside OpenStack to prevent these failures. 


In the case of failures that are notified by API errors without any preceding assertion check failure (the second category in Table~\ref{tab:all_components_all_latencies_table_last_activation}), the latency of the API errors was relatively small, less than one second in the majority of cases. 
Nevertheless, there were few cases with an API error latency higher than one minute. In particular, these cases happened when bugs were injected in Nova, but the API error was raised by a different sub-system (Cinder). In these cases, the high latency was caused by the propagation of the bug's effects across different API calls. These cases are further discussed in \S{}~\ref{subsec:rq3}.

\begin{table}
    \begin{center}
        \footnotesize
        \caption{Statistics on API Error latency.}
        \label{tab:all_components_all_latencies_table_last_activation}
        \vspace{-3mm}
        
        \begin{tabular}{ >{\centering}m{0.8in} |  >{\centering}m{0.4in} | >{\centering}m{0.45in} | >{\centering}m{0.45in} | >{\centering\arraybackslash}m{0.45in}  | }
            \cline{2-5}
            & 
            \textbf{Subsys.}    & \textbf{Avg [s]} & \textbf{50$^{th}$ \%ile  [s]} & \textbf{90$^{th}$ \%ile  [s]} \\
            \hline
            
            
            

            \multicolumn{1}{ |c| }{
            \multirow{3}{*}{\textbf{\shortstack{API Errors after\\ an Assertion\\ failure}}} }
            & Nova & 152.25  & 168.34 & 191.60  \\
            \cline{2-5}
            \multicolumn{1}{ |c|  }{}
            & Cinder & 74.52 & 93.00 & 110.00 \\
            \cline{2-5}
            \multicolumn{1}{ |c|  }{}
            & Neutron & 144.72 & 166.00 & 263.60 \\
            \hline
            




            \multicolumn{1}{ |c| }{
            \multirow{3}{*}{\textbf{\shortstack{API Errors\\ only}}} }
            & Nova & 3.73 & 0.21 & 0.55 \\
            \cline{2-5}
            \multicolumn{1}{ |c|  }{}
            & Cinder & 0.30 & 0.01 & 1.00 \\
            \cline{2-5}
            \multicolumn{1}{ |c|  }{}
            & Neutron & 0.30 & 0.01 & 1.00 \\
            \hline

        \end{tabular}
        \vspace{-0.5cm}
    \end{center}
\end{table}


\subsection{Is OpenStack able to log failures?}
\label{subsec:rq2}

Since failures can be notified to the end-user with a long delay, or even not at all, it becomes important for system operators to get additional information to troubleshoot these failures. In particular, we here consider log messages produced by OpenStack sub-systems.




            
            


We computed the percentage (\textbf{logging coverage}) of failed tests which produced at least one high-severity log message (see also \S{}~\ref{subsec:failure_analysis}). Table \ref{tab:all_components_logged_not_logged_failed_tests} provides the logging coverage for different subsets of failures, by dividing them with respect to the injected subsystem and to the type of failure. 
From these results, we can see that OpenStack logged at least one high-severity message (i.e., with severity level \textit{ERROR} or \textit{CRITICAL}) in most of the cases. The Cinder subsystem shows the best results since logging covered almost all of the failures caused by fault injection. However, in the case of Nova and Neutron, logs missed some of the failures. In particular, the failures without API errors (i.e., \emph{Assertion Failure only}) exhibited the lowest logging coverage. This behavior can be problematic for recovery and troubleshooting since the failures without API errors lack an explicit error notification. These failures are also the ones in need of complementary sources of information, such as logs.

To identify opportunities to improve logging in OpenStack, we analyzed the failures without any high-severity log across, with respect to the bug types injected in these tests. We found that \textit{MISSING FUNCTION CALL} and \textit{WRONG RETURN VALUE} represent the 70.7\% of the bug types that lead to non-logged failures (43.9\% and 26.8 \%, respectively).    
%
%
The \textit{WRONG RETURN VALUE} faults are the easiest opportunity for improving logging and failure detection since the callers of a function could perform additional checks on the returned value and record anomalies in the logs.
For example, one of the injected bugs introduced a \textit{WRONG RETURN VALUE} in calls to a database API called by the Nova sub-system to update the information linked to a new instance. The bug forced the function to return a \emph{None} instance object. The bug caused an assertion check failure, but OpenStack did not log any high-severity message. By manually analyzing the logs, we could only find one suspicious message with the only \textit{WARNING} severity and with little information about the problem, as this message was not related to database management.

The non-logged failures caused by a \textit{MISSING FUNCTION CALL} emphasize the need for redundant end-to-end checks to identify inconsistencies in the state of the virtual resources. For example, in one of these experiments, we injected a \textit{MISSING FUNCTION CALL} in the \textit{LibvirtDriver} class in the Nova subsystem, which allows OpenStack to interact with the \textit{libvirt} virtualization APIs \cite{libvirt}. Because of the injected bug, the Nova driver omits to attach a volume to an instance, but the Nova sub-system does not perform checks that the volume is indeed attached to the instance. This kind of end-to-end checks could be introduced at the service API interface of OpenStack (e.g., in \emph{nova-api}) to test the availability of the virtual resources at the end of API service invocations (e.g., by pinging them).



\begin{table}
    \begin{center}
        \footnotesize
        \caption{Logging coverage of high-severity log messages.}
        \label{tab:all_components_logged_not_logged_failed_tests}
        \vspace{-3mm}
        
        \begin{tabular}{ >{\centering}m{0.75in} | >{\centering}m{0.7in} | >{\centering}m{0.7in} | >{\centering\arraybackslash}m{0.6in}  | }
            
            \cline{2-4}
            & \multicolumn{3}{c|}{\textbf{Logging coverage}} \\
            
            \hline
            \multicolumn{1}{|c|}{\textbf{Subsystem}}    & \textbf{API Errors\\only} & \textbf{Assertion\\ failure only} & \textbf{Assertion failure and API Errors} \\
            \hline
            
            \multicolumn{1}{|c|}{Nova} & 90.32\% & 80.77\% & 82,56\% \\
            \hline
            \multicolumn{1}{|c|}{Cinder} & 100\% & 95,65\% & 100\% \\
            \hline
             \multicolumn{1}{|c|}{Neutron} & 98.67\% & 66.67\% & 95\% \\
            \hline
        \end{tabular}
        \vspace{-0.6cm}
    \end{center}
\end{table}

\subsection{Do failures propagate across OpenStack?}
\label{subsec:rq3}


We analyze failure propagation across sub-systems, to identify more opportunities to reduce their severity. We consider failures of both the ``faulty'' and the ``fault-free'' rounds, respectively (\figurename{}~\ref{fig:workflow}).

In the faulty round, we are interested in whether the injected bug impacted on sub-systems beyond the injected one. To this aim, we divide the API errors with respect to the API that raised the error (e.g., an API exposed by Nova, Neutron, or Cinder). Similarly, we divide the assertion check failures with respect to the sub-system that manages the virtual resource checked by the assertion. There is a \textbf{spatial} fault propagation across the components if an injection on a sub-system (say, Nova) causes an assertion check failure or an API error on a different sub-system (say, Cinder or Neutron).

\begin{figure*}[!ht]
    \begin{centering}
    \begin{subfigure}{0.40\textwidth}
        \includegraphics[width=\linewidth]{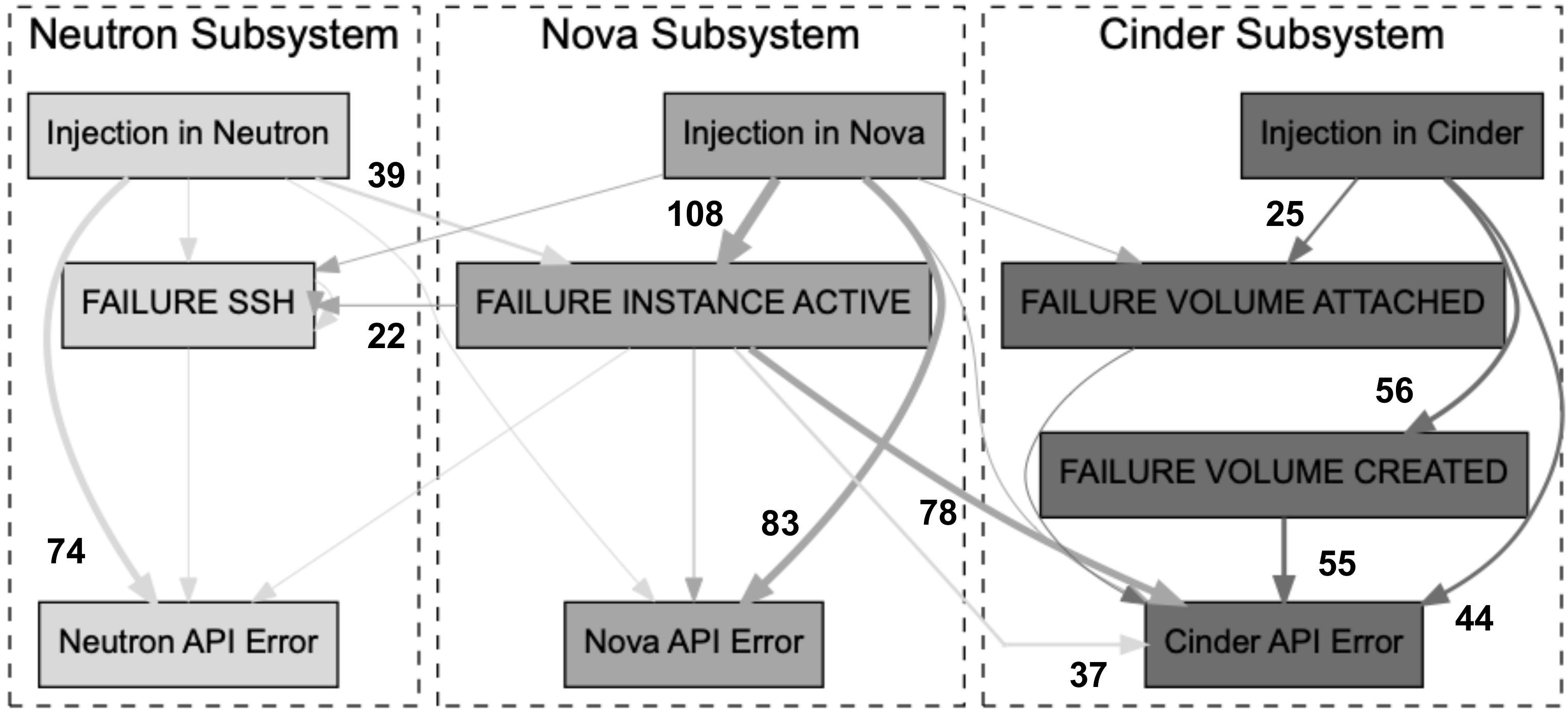}
        \caption{During faulty round.}
        \label{fig:openstack_spatial_propagation_round1}
    \end{subfigure}
    \qquad
    \begin{subfigure}{0.40\textwidth}
        \includegraphics[width=\linewidth]{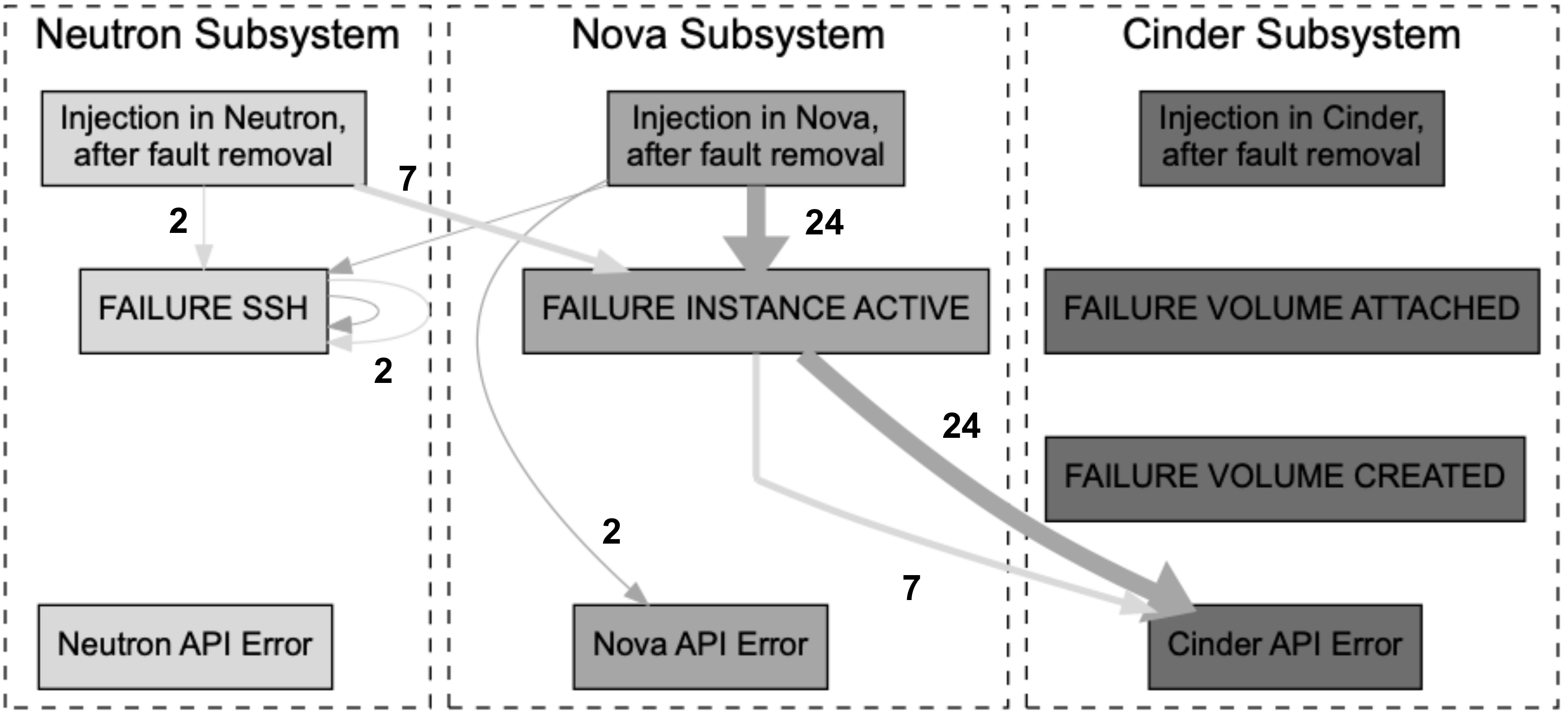}
        \caption{After removing the injected fault (fault-free round).}
        \label{fig:openstack_spatial_propagation_round2}
    \end{subfigure}

    \end{centering}
    \vspace*{-4mm} 
    \caption{Fault propagation during fault injection tests.}
\vspace{-5mm}
\end{figure*}

\figurename{}~\ref{fig:openstack_spatial_propagation_round1} shows a graph with of events occurred during the faulty round of the tests with a failure. The nodes on the top of the graph represent the sub-systems where bugs were injected; the nodes on the middle represent assertion check failures; the nodes on the bottom represent API errors. The edges that originate from the nodes on the top represent the number of injections that were followed by an assertion check failure or an API error. Moreover, the edges between the middle and the bottom nodes represent the number of tests where an assertion check failure was followed by an API error. The most numerous cases are emphasized with proportionally thicker edges and annotated with the number of occurrences. We used different shades to differentiate the cases with respect to the injected sub-system.

            


The failures exhibited a propagation across OpenStack services in a significant amount of cases (37.5\% of the failures). 
In many cases, the propagation initiated from an injection in Nova, which caused a failure at activating a new instance; as discussed in the previous subsections, the unavailability of the instance was detected in a later stage, such as when the user attaches a volume to the instance using the Cinder API. 
Even worse, there are some cases of propagation from Neutron across Nova and Cinder.
These failures represent a severe issue for fault containment since an injection in Neutron not only caused a failure of their APIs but also impacted on virtual resources that were not managed by these sub-systems. Therefore, the failures are not necessarily limited to the virtual resources managed by the sub-system invoked at the time of the failure, but also to other related virtual resources. Therefore, end-to-end checks on API invocations should also include resources that are indirectly related to the API (such as, checking the availability of an instance after attaching a volume).
For as concerns Cinder, instead, there are no cases of error propagation from this sub-system across Nova and Neutron.


We further analyze the propagation of failures by considering what happens during the fault-free round of execution. The fault-free round invokes the service APIs after the buggy execution path is disabled as dead code. Moreover, the fault-free round executes on new virtual resources (i.e., instances, networks, routers, etc., are created from scratch). 
Therefore, it is reasonable to expect (and it is indeed the case) that the fault-free round executes without experiencing any failure. However, we still observe a subset of failures (7.5\%) that propagate their effects to the fault-free round. These failures must be considered critical, since they are affecting service requests that are supposed to be independent but are still exposed to \textbf{temporal} failure propagation through shared state and resources. 
We remark that the failures in the fault-free round are caused by the injection in the faulty round. Indeed, we assured that previous injections do not impact on the subsequent experiments by restoring all the persistent state of OpenStack before every experiment.

\figurename{}~\ref{fig:openstack_spatial_propagation_round2} shows the propagation graph for the fault-free round. The most cases, the Nova sub-system was unable to create new instances, even after the injected bug is removed from Nova. A similar persistent issue happens for a subset of failures caused by injections in Neutron. These sub-systems both manage a relational database which holds information on the virtual instances and networks, and we found that the persistent issues are solved only after that the databases are reverted to the state before fault injection. This recovery action can be very costly since it can take a significant amount of time, during which the cloud infrastructure may become unavailable. For this reason, we remark the need for detecting failures as soon as they occur, such as using end-to-end checks at the end of service API calls. Such detection would support quicker recovery actions, such as to revert the database changes performed by an individual transaction.

%
%
%
%




\subsection{Discussion and lessons learned}

The experimental analysis pointed out that software bugs often cause erratic behavior of the cloud management system, hindering detection and recovery of failures. We found failures that were notified to the user only after a long delay when it is more difficult to trace back the root cause of the failure, and recovery actions are more costly (e.g., reverting the database); or, the failures were not notified at all. 
Moreover, our analysis suggests the following practical strategies to mitigate these failures. 


\noindent
$\rhd$ \textbf{Need for deeper run-time verification of virtual resources.} 
Fault injections pointed out OpenStack APIs that leaked resources on failures, or left them in an inconsistent state, due to missing or incorrect error handlers.
For example, the \emph{server-create} API failed without creating a new VM, but it did not deallocate virtual resources (e.g., instances in ``dead'' state, unused virtual NICs) created before the failure. 
These failures can be prevented through fault injection. Moreover, residual faults should be detected and handled by means of run-time verification strategies, which perform redundant, end-to-end checks after a service API call, to assert whether the virtual resources are in the expected state. For example, these checks can be specified using temporal logic and synthesized in a run-time monitor \cite{delgado2004taxonomy,chen2007mop,zhou2014runtime,rabiser2017comparison}, e.g., a logical predicate for a traditional OS can assert that a thread suspended on a semaphore leads to the activation of another thread \cite{arlat2002dependability}. In the context of cloud management, the predicates should test at run-time the availability of virtual resources (e.g., volumes and connectivity), similarly to our assertion checks (\tablename{}~\ref{tab:assertion_and_description}).

\noindent
$\rhd$ \textbf{Increasing the logging coverage.} 
The logging mechanisms in OpenStack reported high-severity error messages for many of the failures. However, there were failures with late or no API errors that would benefit from logs to diagnose the failure, but such logs were missing. In particular, fault injection identified function call sites in OpenStack where the injected wrong return values were ignored by the caller. These cases are opportunities for developers to add logging statements and to improve the coverage of logs (e.g., by checking the outputs produced by the faulty function calls). Moreover, the logs can be complemented with the run-time verification checks.

\noindent
$\rhd$ \textbf{Preventing corruptions of persistent data and shared state.} 
The experiments showed that undetected failures can propagate across several virtual resources and sub-systems. Moreover, we found that these propagated failures can impact on shared state and persistent data (such as databases), causing permanent issues. Fault injection identified failures that were detected much later after their initial occurrence (i.e., with high API error latency, or no API errors at all). In these cases, it is very difficult for operators to diagnose which parts of the system have been corrupted, thus increasing the cost of recovery. Therefore, in addition to timely failure detection (using deeper run-time verification techniques, as discussed above), it becomes important to address the corruptions as soon as the failure is detected, since the scope of recovery actions can be smaller (i.e., the impact of the failure is limited specific resources involved by the failed service API call). One potential direction of research is on selectively undoing recent changes to the shared state and persistent data of the cloud management system \cite{weber2012automatic,satyal2017rollback}.

\subsection{Threats to validity}
\label{subsec:threats_validity}

The injection of software bugs is still a challenging and open research problem. We addressed this issue by using code mutations to generate realistic run-time errors. This technique is widespread in the field of mutation testing \cite{jia2011survey,just2014mutants,papadakis2018mutation,papadakis2019mutation} to devise test cases; moreover, it is also commonly adopted by studies on software dependability \cite{chillarege1996:generation-error-set,voas1997:predicting,ng2001design,duraes2006emulation,giuffrida2013edfi} and on assessing bug finding tools \cite{dolan2016lava,roy2018bug}. In our context, bug injection is meant to anticipate the potential consequences of bugs on service availability and resource integrity. 
To strengthen the connection between the real and the experimental failures, we based our selection of code mutations on past software bugs in OpenStack. 
The injected bug types were consistent with code mutations typically adopted for mutation testing and fault injection (e.g., the omission of statements). Moreover, the analysis of OpenStack bugs gave us insights on where to apply the injections (e.g., on method calls for controlling Nova, for performing SQL queries, etc.).  
Even if some categories of failures may have been over- or under-represented (e.g., the percentages for failures that were not detected or that propagated), our goal is to point out the existence of potential, critical classes of failures, despite possible errors in the estimates of the percentages. In our experiments, these classes were large enough to be considered a threat to cloud management platforms.

%% file: related.tex

\noindent
$\rhd$ \textbf{Analysis of bugs and failures of cloud systems.} 
Previous studies on the nature of outages in cloud systems analyzed the failure symptoms reported by users and developers, and the bugs in the source code that caused these failures.
\noindent
Among these studies Li et al. \cite{li2018empirical} analyzed failures of Amazon Elastic Compute Cloud APIs and other cloud platforms, by looking at failure reports on discussion forums of these platforms. They proposed a new taxonomy to categorize both failures (content, late timing, halt, and erratic failures) and bugs (development, interaction, and resource faults). One of the major findings is that the majority of the failures exhibit misleading content and erratic behavior. Moreover, the work emphasizes the need for counteracting ``development faults'' (i.e., bugs) through ``semantic checks of reasonableness'' of the data returned by the cloud system. Musavi et al. \cite{musavi2016experience} focused on API issues in the OpenStack project, by looking at the history of source-code revisions and bug-fixes of the project. They found that most of the changes to API are meant to fix API issues and that most of the issues are due to ``programming faults''.
Gunawi et al. analyzed outage failures of cloud services \cite{gunawi2016does}, by inspecting headline news and public post-mortem reports, pointing out that software bugs are one of the major causes of the failures. In a subsequent study, Gunawi et al. analyzed software bugs of popular open-source cloud systems \cite{gunawi2014bugs}, by inspecting their bug repositories. 
The bug study pointed out the existence of many ``killer bugs'' that are able to cause cascades of failures in subtle ways across multiple nodes or entire clusters; and that software bugs exhibit a large variety, where ``logic-specific'' bugs represent the most frequent class. Most importantly, the study remarks that cloud systems tend to favor availability over correctness: that is, the systems attempt to continue running despite the bugs cause data inconsistencies, corruptions, or low-level failures are detected, in order to avoid that users could perceive outages, but putting at risk the correctness of the service.
\noindent
These studies give insights into the nature of failures in cloud systems and point out that software bugs are a predominant cause of failures. While these studies rely on evidence that was collected ``after the fact'' (e.g., the failure symptoms reported by the users), we analyze failures in a controlled environment through fault injection, to get more detailed information on the impact on the integrity of virtual resources, error logs, failure propagation, and API errors.


\vspace{2pt}
\noindent
$\rhd$ \textbf{Fault injection in cloud systems.} The fault injection is widely used for evaluating fault-tolerant cloud computing systems. 
Well-known solutions in this field include \emph{Fate} \cite{Gunawi2011a} and its successor \emph{PreFail} \cite{Joshi2011b} for testing cloud software (such as Cassandra, ZooKeeper, and HDFS) against faults from the environment, by emulating at API level the unavailability of network and storage resources, and crashes of remote processes. Similarly, Ju et al. \cite{Ju2013a} and \emph{ChaosMonkey} \cite{chaos_monkey} test the resilience of cloud infrastructures by injecting crashes (e.g., by killing VMs or service processes), network partitions (by disabling communication between two subnets), and network traffic latency and losses.
Other fault models for fault injection include hardware-induced CPU and memory corruptions, and resource leaks (e.g., induced by misbehaving guests). \emph{CloudVal} \cite{Pham2011} and Cerveira et al. \cite{cerveira2015recovery} applied these fault models to test the isolation among hypervisors and VMs. Pham et al. \cite{pham2017failure} applied fault injection on OpenStack to create signatures of the failures, in order to support problem diagnosis when the same failures happen in production. 
\noindent
The fault model is the main difference that distinguishes our work from previous studies. Most of them assess software robustness with respect to \emph{external} events (e.g., a faulty CPU, disk or network). 
In other studies, fault injection has been simulating software failures through process crashes and API errors, but this is a simplistic form of software bugs, which can cause generate more subtle effects (such as incorrect logic and data corruptions, as pointed out by bug studies). In this work, we injected \emph{software bugs} inside components by mutating their source code, to deliberately force their failure, and to assess what happens in the worst case that a bug eludes the QA process and gets into the deployed software.

\noindent
We remark that previous work on mutation testing \cite{jia2010analysis} also adopted code mutation, but with a different perspective than ours, since we leverage mutations for evaluating software fault tolerance. Our work contributes to this research field by showing new forms of analysis based on the injection of software faults (fail-stop behavior, logging, failure-propagation). The same approach is also suitable to other systems of similar size and complexity of OpenStack (e.g., where the need for coordination among large subsystems raises the risk for non-fail-stop behavior and failure propagation).

%% file: artifacts.tex

We release the following artifacts to support future research on mitigating the impact of software bugs: \textit{(i)} the analysis of OpenStack bug reports (\url{https://doi.org/10.6084/m9.figshare.7731629}), \textit{(ii)} raw logs produced by the experiments (\url{https://doi.org/10.6084/m9.figshare.7732268}), and \textit{(iii)} tools for reproducing our experimental environment in a virtual machine (\url{https://doi.org/10.6084/m9.figshare.8242877}).


%% file: conclusion.tex

In this work, we proposed a methodology to assess the severity of failures caused by software bugs, through the deliberate injection of software bugs. We applied this methodology in the context of the  OpenStack cloud management system. The experiments pointed out that the behavior of OpenStack under failure is not amenable to automated detection and recovery. In particular, the system often exhibits a \emph{non-fail-stop} behavior, in which it continues to execute despite inconsistencies in the state of the virtual resources, without notifying the user about the failure, and without producing logs for aiding system operators. Moreover, we found that the failures can spread across several sub-systems before being notified and that they can cause persistent effects that are difficult to recover. Finally, we point out areas for future research to mitigate these issues, including run-time verification techniques to detect subtle failures in a more timely fashion and to prevent persistent corruptions.